\documentclass[%
 aps,
 prb,%
 amsmath,amssymb,showpacs,
 reprint,%
]{revtex4-1}

\usepackage{graphicx}
\usepackage{dcolumn}
\usepackage{bm}

\begin{document}


\title[Sample title]{Anomalous dissipation mechanism and Hall quantization limit\\
in polycrystalline graphene grown by chemical vapor deposition}

\author{F. Lafont$^1$, R. Ribeiro-Palau$^1$, Z. Han$^2$, A. Cresti$^3$, A. Delvall\'ee$^1$, A. W. Cummings$^4$, S. Roche$^{4,5}$, V. Bouchiat$^2$, S. Ducourtieux$^1$, F. Schopfer$^1$ and W. Poirier$^1$}
\affiliation{$^1$LNE - Laboratoire National de M\'{e}trologie et d'Essais, 78197 Trappes, France}
\affiliation{$^2$Institut N\'eel, Centre National de la Recherche Scientifique - Universit\'e Joseph Fourier - Grenoble Institue of Technology, 38042 Grenoble, France}
\affiliation{$^3$Institute of Microelectronics, Electromagnetism, and Photonics- Laboratoire d'Hyperfr\'equences et de Caract\'erisation (UMR5130), Grenoble Institute of Technology MINATEC, 38016 Grenoble, France}
\affiliation{$^4$Institut Català de Nanosci\`encia i Nanotecnologia - Autonomous University of Barcelona, 08193 Bellaterra, Spain}
\affiliation{$^5$Instituci\`o Catalana de Recerca i Estudis Avan\c{c}ats, 08010 Barcelona, Spain}
\date{\today}

\begin{abstract}
We report on the observation of strong backscattering of charge carriers in the quantum Hall regime of polycrystalline graphene, grown by chemical vapor deposition, which alters the accuracy of the Hall resistance quantization. The temperature and magnetic field dependence of the longitudinal conductance exhibits unexpectedly smooth power law behaviors, which are incompatible with a description in terms of variable range hopping or thermal activation, but rather suggest the existence of extended or poorly localized states at energies between Landau levels. Such states could be caused by the high density of line defects (grain boundaries and wrinkles) that cross the Hall bars, as revealed by structural characterizations. Numerical calculations confirm that quasi-1D extended non-chiral states can form along such line defects and short-circuit the Hall bar chiral edge states.
\end{abstract}

\maketitle
\section{\label{sec:level1}Introduction}
One manifestation of the Dirac physics in graphene is a quantum Hall effect (QHE) \cite{Novoselov2005,Zhang2005} with an energy spectrum quantized in Landau levels (LLs) at energies $E_n=\pm v_{\mathrm{F}}\sqrt{2\hbar n e B}$, with a $4eB/h$ degeneracy (valley and spin) \cite{Goerbig2011} and a sequence of Hall resistance plateaus at $R_{\mathrm{H}}=\pm R_{\mathrm{K}}/[4(n+1/2)]$, where $n\geqslant 0$ and $R_{\mathrm{K}}\equiv h/e^2$. The QHE at LLs filling factor $\nu=\pm2$ ($\nu=n_\mathrm{s}h/eB$, where $n_\mathrm{s}$ is the carrier density) is very robust and can even survive at room temperature \cite{Novoselov2007}. This comes from an energy spacing $\Delta E(B)\approx 35\sqrt{B[\mathrm{T}]}~\mathrm{meV}$ between the first two degenerated LLs, which is larger than in GaAs ($\approx1.7B[\mathrm{T}]~\mathrm{meV}$), for accessible magnetic fields. This opens the door for a $10^{-9}$-accurate quantum resistance standard in graphene, surpassing the usual GaAs-based one, in operating at lower magnetic fields ($B\leq$ 4 T), higher temperature ($T\geq$ 4 K) and higher measurement current ($I\geq100~\mu$A) \cite{Poirier2010}. From previous investigations of the QHE in graphene \cite{Giesberg2009,Tzalenchuk2010,Guignard2012,Wosczczyna2012}, it was concluded that achieving this goal requires at least the production of a large area graphene monolayer ($\sim 10~000~\mathrm{\mu m^{2}}$) of high carrier mobility $\mu > 10~000~\mathrm{cm^{2}V^{-1}s^{-1}}$ (assuming $\mu B\gg1$ stays a relevant quantization criterion \cite{Schopfer2012}) and homogeneous low carrier density ($n_{\mathrm{s}}<2\times 10^{11}\mathrm{cm^{-2}}$). However, the question arises whether some defects, specific to each source of graphene, can jeopardize the quantization accuracy. It was thereby shown, using exfoliated graphene, that the presence of high density of charged impurities in the substrate on which graphene lies can limit the robustness of the Hall resistance quantization by a reduction of the breakdown current of the QHE \cite{Guignard2012}.
\begin{figure}[h!]
\begin{center}
\includegraphics[width=8.5cm]{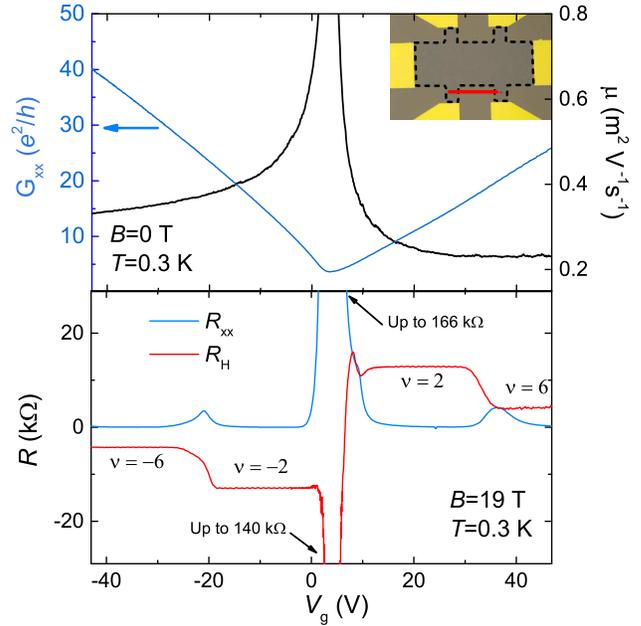}
\caption{(a) Longitudinal conductance and carrier mobility vs. $V_g$ and (b) $R_\mathrm{H}$ and $R_\mathrm{xx}$ vs. $V_g$ for sample S1. Insert in (a): Hall bar optical image. The length scale (red segment) between voltage   terminals is $200~\mathrm{\mu m}$ and equal to the Hall bar width.}\label{fig1}
\end{center}
\end{figure}
Although the quantization of $R_{\mathrm{H}}$ was measured with an uncertainty of $9\times 10^{-11}$ in a large $35\times 160~\mu \mathrm{m}^2$ sample made of graphene grown by sublimation of silicon from silicon carbide, at 14 T and $0.3~\textrm{K}$ \cite{Janssen2011}, it was recently demonstrated, both experimentally\cite{Chua2014} and theoretically\cite{Lofwander2013}, that bilayer stripes forming along the silicon-carbide edge steps during the growth and crossing the Hall bar, can short-circuit the edge states and strongly alter the Hall quantization.

Growth based on chemical vapor deposition (CVD) appears to be a promising route to produce large-area graphene with high mobility \cite{Petrone2012,Cummings2014}. The QHE is now commonly observed in such graphene. However, in a $\mathrm{7\times7~mm^2}$ sample, $R_{\mathrm{H}}$ at $\nu =2$ was found to deviate from $R_\mathrm{K}/2$ by more than $10^{-2}$, while the longitudinal resistance per square reached $R_\mathrm{xx} =200~\Omega$ \cite{Shen2011}, which is the mark of a high dissipation, still unexplained. In comparison, a GaAs-based quantum resistance standard satisfies $R_\mathrm{xx} < 100~\mathrm{\mu \Omega}$. This highlights the need for exploration of the precise electronic transport mechanisms at work in CVD graphene.

In this paper, we investigate the QHE in large Hall bars made of polycrystalline CVD graphene. We observe a strong dissipation characterized by an unexpected power law dependence of the conductance with \emph{T}, \emph{B}, and \emph{I}, which reveals an unconventional carrier backscattering mechanism. Structural characterizations bring out line defects crossing the devices, such as grain boundaries (GBs) or wrinkles naturally existing in polycrystalline CVD graphene. While some works exist at $B=0$ T \cite{Tsen2012, Tuan2013, Yazyev2010, Zhu2012, Pereira2010}, the impact on transport of these line defects has been hardly investigated, to our knowledge, in the QHE regime \cite{Jauregui2011, Ni2012, Calado2014}. With the support of numerical simulations we highlight their paramount role in limiting the Hall quantization.
\section{\label{sec:level1}Sample fabrication}
Large scale graphene films were grown on Cu foils by standard CVD method. In this process, gaseous methane [2 sccm (sccm denotes standard cubic centimeter per minute at STP)] and hydrogen (70 sccm) precursors were introduced into a quartz tube reactor heated at 1000 $^{\circ}$C for 40 min under a total pressure of 1 mbar. After cooling, graphene was transferred onto a Si wafer with 285 nm thick SiO$_2$ layer, by etching the underneath Cu, using 0.1 g/ml $\mathrm{(NH_4)_2S_2O_8}$ solution \cite{Han2014}. The Hall bar samples studied in the paper were fabricated by optical lithography, oxygen plasma etching and contacted with  Ti/Au (5 nm/60 nm) electrodes. Both samples (S1 and S2) were grown and transferred in the same process. Sample S1 was measured as fabricated while sample S2 was annealed at $110\, ^{\circ}$C in a H$_2$/Ar atmosphere during 10 hours. Hall bars dimensions are $ 200 \times 400~\mathrm{\mu m^{2}}$ (inset of Fig. \ref{fig1}(a)). Main magneto-transport results concern sample S1, results in sample S2 are used to illustrate reproducibility and sample independence. For this, unless specified, results and discussions concern sample S1.
\section{\label{sec:level1}Results and discussion}
\subsection{\label{sec:level2} Conductance laws}
Figure \ref{fig1}(a) shows the conductance at zero magnetic field deduced from the resistance per square $G_{xx}=1/R_{xx}$, $G_{xx}$ as a function of the gate voltage $V_g$ at $0.3~\mathrm{K}$. The charge neutrality point (CNP) is positioned at $V_g=3.5~\mathrm{V} $, which indicates a residual hole density of $\mathrm{\sim2.6\times 10^{11}cm^{-2}}$, assuming a $\mathrm{SiO_2}$/Si back-gate efficiency  of $7\times10^{10}~\mathrm{cm^{-2}/V}$. At high carrier density ($\sim1\times 10^{12}\mathrm{cm^{-2}}$), the hole (electron) mobility is $\sim 3100~\mathrm{{cm}^{2}V^{-1}s^{-1}}$ ($\sim 2300~\mathrm{{cm}^{2}V^{-1}s^{-1}}$). The electron phase coherence length $L_\mathrm{\phi}$, the inter-valley scattering length $L_\mathrm{iv}$, and the intra-valley scattering length are $\mathrm{0.9~\mu m}$, $\mathrm{0.3~\mu m}$ and $\mathrm{0.1~\mu m}$, respectively, as deduced from the measurement (see Appendix A) of the weak localization correction to the conductance at $0.3~\mathrm{K}$ \cite{McCannWL2006}. The lower value of $L_\mathrm{iv}$ compared to $L_\mathrm{\phi}$ indicates the presence of a significant concentration of short-range scatterers.
\begin{figure}[h!]
\begin{center}
\includegraphics[width=8.5cm]{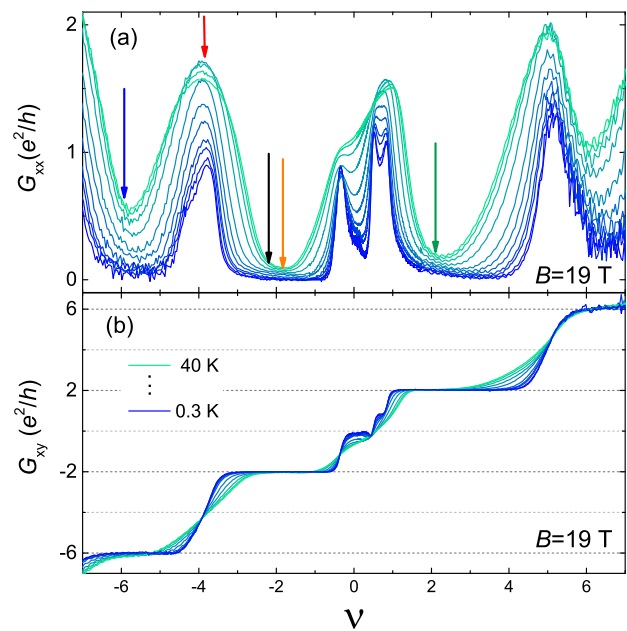}
\caption{(a) $G_\mathrm{xx}$ and (b) $G_\mathrm{xy}$ vs. $\nu$ for \emph{T} between 0.3 K and 40 K at 19 T, obtained in sample S1. Temperature color code apply for both figures. Arrows in (a) indicate the values of $\nu$ at which measurements in Fig. 3(a) are performed.}\label{fig2}
\end{center}
\end{figure}

The Hall resistance, $R_{\mathrm{H}}$, measured at 0.3 K and 19 T, is reported as a function of $V_g$ in Fig. \ref{fig1}(b). It features well-developed $R_{\mathrm{H}}$ plateaus at values $h/\nu e^2$ for $\nu=\pm2,\pm6$, which coincide with the minima of the longitudinal resistance per square $R_\mathrm{xx}$. Close to the CNP, additional high resistance peaks with $R_\mathrm{H},R_\mathrm{xx}\gg h/e^2$ are observed, corresponding to plateaus with transverse conductance $G_\mathrm{xy}= R_{\mathrm{H}} / (R_{\mathrm{H}}^{2} + R_{\mathrm{xx}}^{2})$ around $0$ and $e^2/h$ in Fig. \ref{fig2}(b). These plateaus are accompanied by minima of the longitudinal conductance per square $G_\mathrm{xx}=R_{\mathrm{xx}} / (R_{\mathrm{H}}^{2} + R_{\mathrm{xx}}^{2})$ also located around $\nu=0$ and $\nu=1$, respectively, Fig. \ref{fig2}(a). Such conductance plateaus can be explained by the degeneracy lifting of the $n=0$ LL \cite{Goerbig2011,Kharitonov2012}, which is usually observed in graphene with much higher carrier mobility. We therefore do not exclude the possibility that the carrier mobility inside a monocrystalline grain would be higher than the moderate value calculated from the mean conductance $G_\mathrm{xx}$ averaged over several grains. More extensive analysis of these additional plateaus is beyond the scope of this article.

Although nice plateaus are observed, it turns out that $R_{\mathrm{H}}$ is not well quantized, even on the $\nu=-2$ plateau, deviating from $R_\mathrm{K}/2$ by more than $10^{-2}$ in relative value at a current of $1~\mu$A, while $R_{\mathrm{xx}}$, which reflects the dissipation arising from backscattering between counter-propagating quantum Hall edge states, is
higher than $150~\Omega$. This is unexpected since the quantization of $R_{\mathrm{H}}$ has been measured with uncertainties several orders of magnitude lower in exfoliated samples smaller than ours and with similar carrier mobilities\cite{Giesberg2009,Guignard2012,Wosczczyna2012}. This shows that the transport properties in the QHE regime are very sensitive to the defect-type and that the mobility at $B=0$ T does not constitute a sufficient criteria of quantization.
\begin{figure}[h]
\begin{center}
\includegraphics[width=8.5cm]{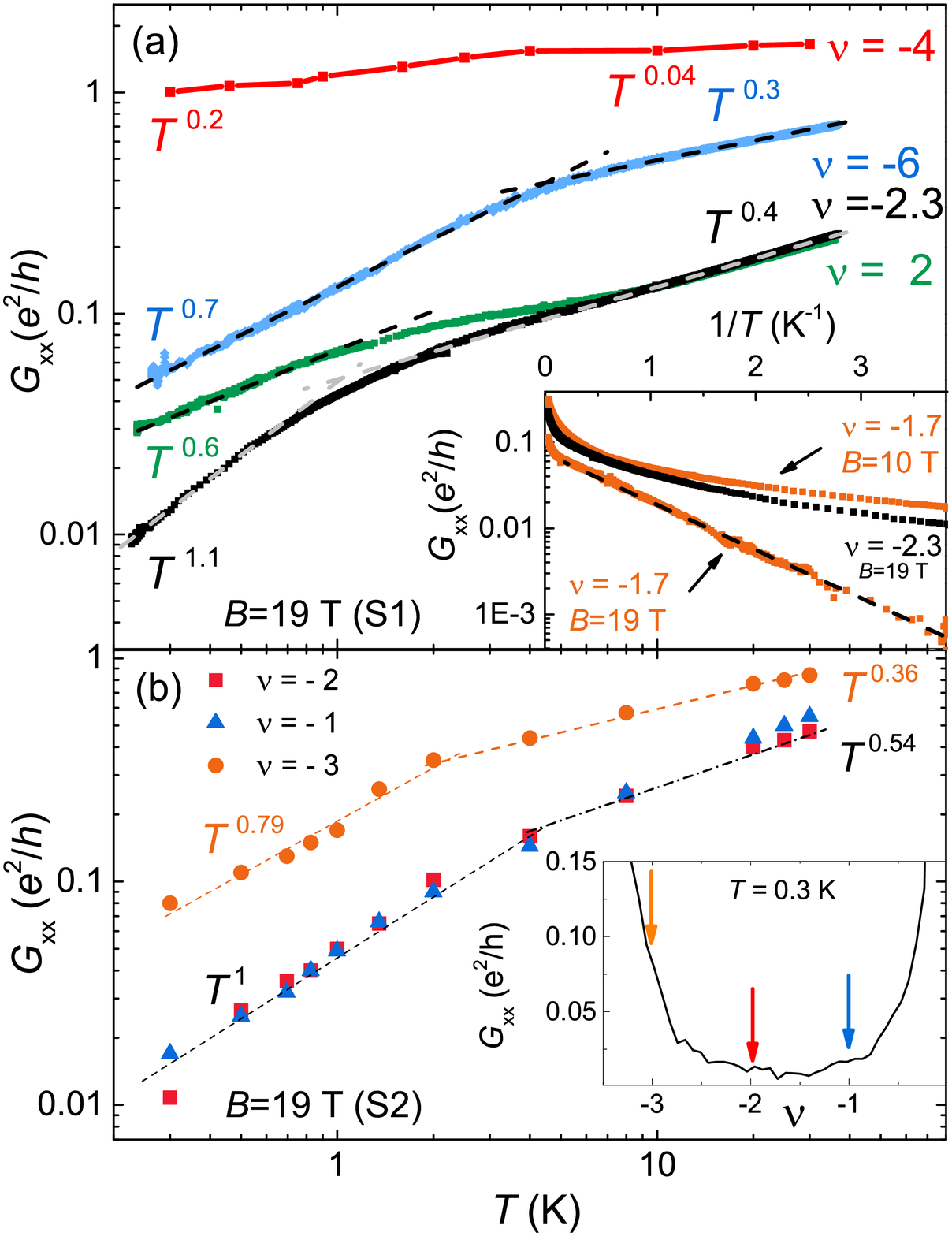}
\caption{(a) $G_\mathrm{xx}$ vs. $T$ in log-log scale at 19 T for S1. Inset: $G_\mathrm{xx}$ in log scale vs. 1/\emph{T} for $\nu=-1.7$ at 19 and 10 T and at $\nu=-2.3$ for comparison. (b) $G_\mathrm{xx}$ vs. $T$ in log-log scale for S2. Inset: $G_\mathrm{xx}$ vs. $\nu$ at 0.3 K, arrows indicate the values of $\nu$ at which measurements are performed.}\label{fig3}
\end{center}
\end{figure}

\begin{figure}[t]
\begin{center}
\includegraphics[width=8cm]{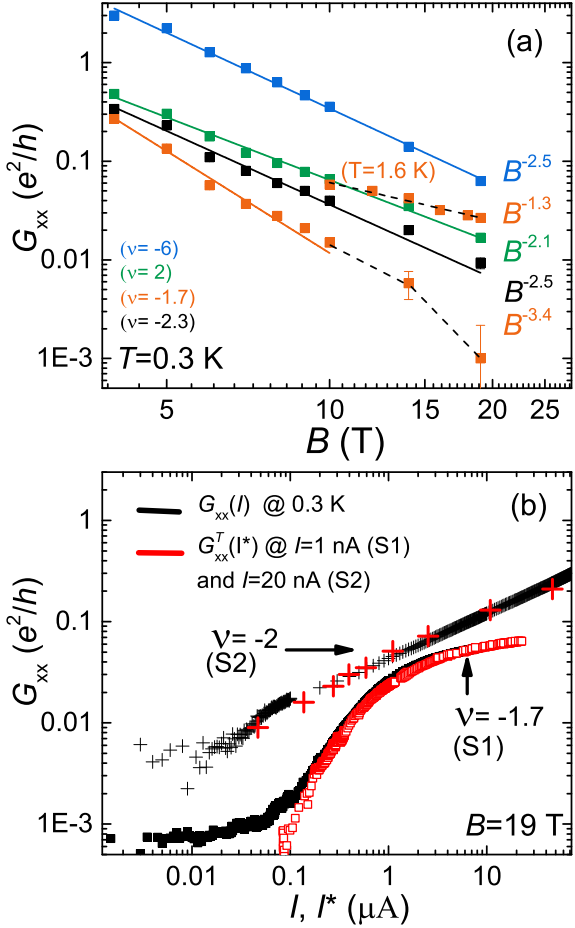}
\caption{(a) $G_\mathrm{xx}$ vs. $B$ in log-log scale at 0.3 K for different filling factors for S1. (b) $G_\mathrm{xx}$ vs. $I$ and $G^T_\mathrm{xx}$ vs. $I^{*}$ in log-log scale for the two samples with $I^{*}[\mathrm{A}]=0.87\times10^{-6}~T[\mathrm{K}]^{1.74}$ for S1 and $I^{*}[\mathrm{A}]=0.6\times10^{-6}~T[\mathrm{K}]^{2.1}$ for S2.}\label{fig4}
\end{center}
\end{figure}

To identify the mechanism responsible for this loss of quantization, we analysed $G_\mathrm{xx}$, known as the quantization parameter \cite{Jeckelmann2001},
 over a large range of $\nu$ values, at several temperatures between 0.3 K and 40 K (see Fig. \ref{fig2}(a)), and at magnetic fields between 5 T and 19 T. Measurements of $R_{\mathrm{H}}$
and $R_{\mathrm{xx}}$ were carried out using a low-frequency AC measurement current of 1 nA, which ensures the absence of current effects, see fig. \ref{fig4}(b). Except for $\nu=-1.7$, where $G_\mathrm{xx}$ reaches its minimum, and at \emph{B}=19 T, it appears for both type of carriers (electrons and holes) that neither $G_\mathrm{xx}(T)$ nor $G_\mathrm{xx}(B)$ (Figs. \ref{fig3}(a) and \ref{fig4}(a), respectively) has an exponential behavior, which would be expected for a dissipation mechanism based on thermal activation to a higher-energy LL or variable range hopping (VRH) through localized states in the bulk. This greatly differs from what has been observed in both exfoliated  \cite{Giesbergact2007,Giesbergvrh2009,Bennaceur2012} and epitaxial graphene \cite{Tzalenchuk2011}. Rather, whatever the quantum Hall state, at $\nu=\pm2$ or $\pm6$, $G_\mathrm{xx}$ follows a power law dependence as a function of
temperature ($G_\mathrm{xx}\propto T^{\alpha}$) and magnetic induction ($G_\mathrm{xx}\propto B^{-\beta}$) with $\alpha\in [0.3,1.1]$ (at 19 T) and $\beta \in [2.1, 3.4]$ (at 0.3 K). The temperature dependence becomes smoother with $\nu$ moving away from the conductance minimum. For $G_\mathrm{xx}(T)$, we can also define two temperature regimes characterized by larger $\alpha$
at lower temperature and a smooth crossover. In a given temperature regime and magnetic field, $\alpha$ slightly varies with $\nu$, away from the LL centers. The same temperature behavior
of $G_\mathrm{xx}$, with similar $\alpha$ values, was observed in sample S2,  Fig. \ref{fig3}(b). In S1, the dependence of $G_\mathrm{xx}$ on $T$ ($B$) becomes smoother with decreasing $B$ (increasing $T$)(Fig. \ref{fig3}(a) and \ref{fig4}(a)), characterized by decreasing values of $\alpha$ ($\beta$). Such behaviors are consistent with a reducing inter-LL energy gap. Interestingly,
the $G_\mathrm{xx}$ power law temperature dependence, observed for $\nu$ corresponding to $G_\mathrm{xx}$ minima, is similar to that observed at $G_\mathrm{xx}$ maxima, where charge transport is known to occur through extended LL states (as shown for $\nu=-4$ in Fig. \ref{fig3}(a)). This suggests the scenario that the strong backscattering observed near $\nu=\pm2$ and $\pm6$ is caused by extended
or poorly localized states existing at energies between LLs.

At $\nu=-1.7$, a fit of $G_\mathrm{xx}(T)$ with an Arhenius law $\propto \exp[-(T_{\mathrm{act}}/T)]$ results in an activation temperature of 2.4 K $\ll\Delta E(B=19~\mathrm{T})/k_\mathrm{B}\sim 1834~\mathrm{K}$ (inset of Fig. 3(a)), suggesting mobility edge energies unexpectedly far from the LL centers and confirming the fragility of the $R_{\mathrm{H}}$ quantization. A fit with a VRH theory including a soft Coulomb gap \cite{Shklovskii1984}, $G_{\rm xx}\propto(1/T)\exp(-(T_0/T)^{1/2})$, is also possible and leads to $T_0=27~\mathrm{K}$ and a high value for the localization length $\xi=Ce^2/(4\pi\epsilon_0\epsilon_r k_\mathrm{B}T_0)$ (with $C\sim6.2$ \cite{Furlan1998}), equal to $\sim 1~\mathrm{\mu m}\gg l_B(19~\mathrm{T})\sim 6~\mathrm{nm}$  \cite{CommentXsi,Bennaceur2012}, which is the mark of poorly localized states in the bulk that can even have a metallic behaviour since $\xi\geq L_\mathrm{\phi}$. Decreasing the magnetic field from 19 T to 10 T, while $\nu$ is fixed at -1.7, results in a transition to a power law temperature dependence [Fig. 3(a)(inset)]. This can be explained once again by the delocalization of states between LLs because of a further increasing increasing $\xi$, and a decreasing inter-LL energy gap.

The analysis of the dependence of $G_\mathrm{xx}$ on the current is also instructive. Near $\nu=-2$, a significant increase of $G_\mathrm{xx}$ starting from currents as low as 100 nA indicates a breakdown current density of the QHE lower than $5\times10^{-3}$ A/m, which is unexpectedly small compared to values measured in epitaxial graphene (up to 43 A/m at 23 T) \cite{Alexander-Webber2013} or in exfoliated graphene 0.5 A/m at 18 T\cite{Wosczczyna2012}. This also suggests the existence of extended states accessible at low electric field. Moreover, Fig. 4(b) shows that a similar current-temperature conversion relationship, $I^{*}\propto T^p$ with $p\sim 2$, exists for both samples S1 and S2. This allows for a good superposition of $G_\mathrm{xx}(I)$ and $G_\mathrm{xx}^T(I^{*})$,
where $G_\mathrm{xx}(T)=G_\mathrm{xx}^T(I^{*})$, on a common current scale at sufficiently high $I$ such that $G_\mathrm{xx}$ is not limited by $T$. A relationship $I\propto T$ is expected in the QHE regime from the VRH mechanism \cite{Furlan1998}, as it has been observed in exfoliated graphene \cite{Bennaceur2012}. On the other hand, $I\propto T^2$ was observed in graphene in the metallic regime, at low magnetic field \cite{Baker2012} or in regime of Schubnikov-de-Haas oscillations \cite{Baker2013} and explained by the coupling of carriers to acoustic phonons. The predicted relationship between the current and the temperature is given by  $I=\sqrt{\sqrt{n_s}A\gamma/R_{\rm xx}(B=0)}T^{2}$ where $n_s$ is the carrier density, $A$ is the sample area and $\gamma=5.36\times 10^{-26}\mathrm{WK^{-4}m}$ is a constant\cite{Kubakaddi2009,Baker2012}. Considering $R_{\rm xx}(B=0)=1.8~\mathrm{k\Omega}$ at $n_s\sim 1\times10^{12}\mathrm{cm^{-2}}$ (hole density corresponding to $\nu=-2$ at \emph{B}=19 T), one calculates $I[\mathrm{A}]\sim1.09\times10^{-6}~T[K]^{2}$ which is in a good agreement with our experimental determination $I^{*}[\mathrm{A}]=0.87\times10^{-6}~T[\mathrm{K}]^{1.74}$ for sample S1 and $I^{*}[\mathrm{A}]=0.6\times10^{-6}~T[\mathrm{K}]^{2.1}$ for sample S2 (see fig. \ref{fig4}(b)). This suggests that we can ascribe our observation of $I\propto T^2$ to the manifestation of a metallic regime, which involves extended or poorly localized states, in a weakened QHE regime.
\begin{figure}[h]
\begin{center}
\includegraphics[width=8.5cm]{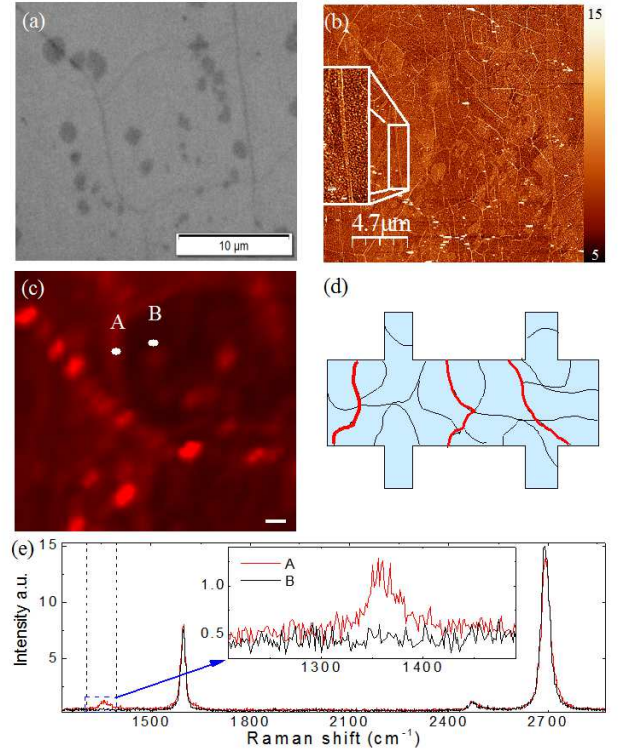}
\caption{(a) Optical and (b) atomic force microscopies. (c) Raman D peak map (scale bar is $1.5~\mathrm{\mu m}$). Figures (a)-(c) concern about the same area of sample S2. (d) Representation of the network of line defects corresponding to short-circuit paths between the sample edges. (e) Raman signal on (A) and away (B) from a wrinkle. Inset: zoom in the D peak zone of the Raman spectra.}\label{fig5}
\end{center}
\end{figure}
\subsection{\label{sec:level2} Structural characterizations}
To better understand our results, complementary structural analyses were performed combining different techniques (Fig. 5). Optical and atomic force microscopy reveal the existence of multilayer patches and a high density and variety of wrinkles. Multilayer patches are known to form locally during CVD growth\cite{Han2014}. Assuming they are located  at the center of the grains, from their pacing we can deduce a typical monocrystalline grain sizes ranging from $\mathrm{1~\mu m}$ to $\mathrm{10~\mu m}$ (GBs were not directly observable with the techniques used).
Given the small size of the patches (Fig. 5(a)) compared to the width of the Hall bars and the ability of carriers to skirt local defects in the QHE regime \cite{Yoshioka1998}, these patches are not expected to cause the observed strong backscattering. In the same way, only large bilayer stripes crossing the Hall bar channel are expected to significantly alter the perfect quantization\cite{Chua2014,Lofwander2013}. Raman spectroscopy in most of the optically clean areas indicates high quality graphene, since no D-peak is observable (Fig. 5(c))\cite{Ferrari2007}. On the other hand, the presence of the D-peak, which confirms the existence of sharp defects, as already revealed by weak localization transport experiments, is measured at locations on most wrinkles. Such a Raman D-peak is the signature of underlying defects such as vacancies or GBs \cite{QingkaiYu2011, Duong2012}. In our samples, wrinkles and GBs are likely to form a continuous network connecting Hall bar edges. Carriers moving from source to drain then cannot avoid crossing some line defects (Fig. 5(d)), which is
expected to impact charge transport.
\begin{figure}[h]
\begin{center}
\includegraphics[width=8.5cm]{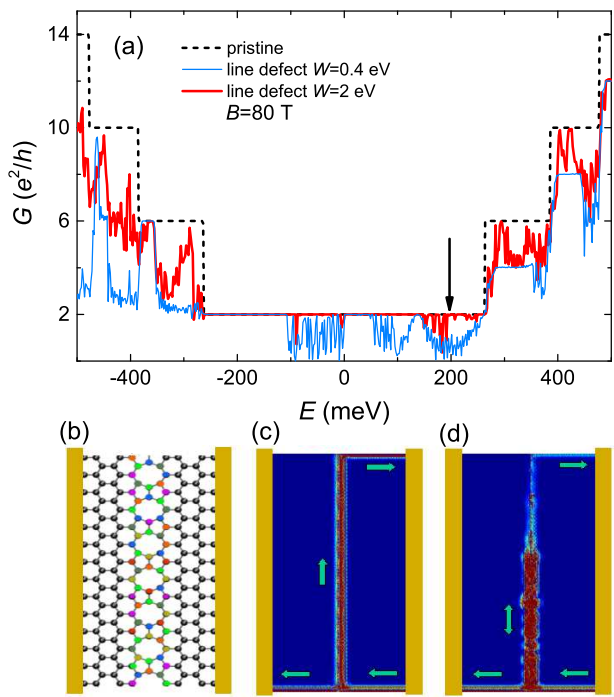}
\caption{(a) Two-terminal magnetoconductance of a pristine aGR, and of aGR with a 8-5 line defect crossing the sample (represented in (b)) including a ramdom disorder potential of $W=0.4$ eV (blue line) and $W=2$ eV (red line). (b) Representation of the  8-5 line defect crossing the aGR. (c) and (d) spatial distribution of the electrons injected from the source contact (to the right) at 200 meV is shown in insets (c) $W=0.4$ eV and (d) $W=2$ eV.}\label{fig6}
\end{center}
\end{figure}
\subsection{\label{sec:level2} Numerical simulations}
To more closely study this impact on the QHE, we performed numerical calculations of the two-terminal conductance of a 200 nm wide armchair graphene ribbon (aGR) crossed by a line of pentagons and octagons \cite{Bahamon2011,Song2012} by using the Green's function approach within the tight-binding framework \cite{Cresti2006}. To simulate a more realistic line defect, a random (Anderson \cite{Anderson1958}) potential with a uniform distribution in the range [-\emph{W}/2,+\emph{W}/2], where \emph{W} is the disorder strength, was introduced on the line defect sites (Fig. 6(b)) to mimic a generic short-range disorder, as the one generated by ad-atoms or vacancies.

In the QHE regime, the calculations reported were performed at \emph{B}=80 T so that $l_B\sim 3~\mathrm{nm}$ is significantly smaller than the ribbon width (in a similar ratio of the experimental $l_B$ to the smallest grain size) and larger than the interatomic distance. For a 100 nm-wide ribbon and \emph{B}=40 T qualitatively very similar results, not shown, were obtained. The calculated conductance almost systematically deviates from the value expected for pristine graphene by up to one spin-degenerated conduction channel [Fig. 6(a)], for weak disorder ($W=0.4~\mathrm{eV}$), significantly larger than what is experimentally observed. The deviation is higher for electrons than for holes, where the asymmetry results from the sublattice symmetry breaking caused by the line defect.
As demonstrated in Fig. 6(c), the deviation of the conductance from the case of pristine graphene is caused by a circulating current along the line defect. An analysis of the energy spectrum shows that counter-propagating states on either side of the line defect can hybridize and form non-chiral quasi-1D extended states \cite{Cummings} able to carry current, which crosslink the opposite sample edges. Acting as a direct short-circuit, such states are responsible for a strong carrier backscattering. Remarkably, higher Anderson disorder reinforces wave-function localization along the line defect and reduces the circulation of current (Fig. 6(d)), which finally improves the Hall conductance quantization. It is also found that, due to the disorder, the deviation of the Hall conductance from pristine quantization reduces with increasing magnetic field and sample width (i.e. the length of the line defect network), both of which enhance the localization. See Appendix B for additional details. Thus, a moderate alteration of the Hall conductance quantization comparable to what is experimentally observed can be reproduced.

Moreover, even though the simulations were run at 0 K, the existence of extended or poorly localized states along the line defect suggests smooth temperature behavior. Localization by strong disorder along the line defect also leads to the possible observation of VRH or thermal activation behavior, characteristic of an Anderson insulator. This is in sound agreement with our experimental observations, since, following the proposed scenario, $G_\mathrm{xx}$ measured at $\nu$ values corresponding to minima should be dominated by the conductance along the line defects, which is much higher than the bulk conductance inside the grains.
Finally, calculations performed for scrolled graphene \cite{Cresti2012} indicate that wrinkles are also expected to alter the Hall conductance quantization in a similar fashion. Recent experimental results also suggest such an impact \cite{Calado2014}.
\section{\label{sec:level1}Conclusion}
To conclude, in polycrystalline CVD graphene characterized by a high density of line defects such as GBs and wrinkles, we highlight an unusual highly dissipative electronic transport in the QHE regime, which reveals the existence of poorly localized states between LLs and manifests itself as a deviation of $R_{\mathrm{H}}$ from the pristine quantization. Numerical simulations confirm that such states can exist along a line defect crossing a Hall bar and yielding strong backscattering between edge states. The impact of line effects turn out to be similar to that of crossing bilayer stripes in graphene grown by sublimation of silicon from silicon carbide\cite{Chua2014}. Further theoretical work, possibly considering Coulomb interactions and Luttinger physics \cite{Fisher1997}, is required to explain the observed temperature, magnetic field and current dependence of $G_\mathrm{xx}$. Our work also motivates the investigation of the QHE in CVD graphene monocrystals, whose size is continuously in progress \cite{Zhou2013}, not only to discern the respective roles of GBs and wrinkles but also to progress towards an operational graphene-based quantum resistance standard. More generally, QHE turns out to be an extremely efficient tool to reveal line defects in 2D materials whose precise
characterization is crucial in view of future applications.
\begin{acknowledgments}
We wish to acknowledge D. Leprat and L. Serkovic for technical support, D. C. Glattli, J.-N. Fuchs, M. O. Goerbig, S. Florens and Th. Champel for fruitful discussions. This research has received funding from the Agence national de la Recherche (ANR) , Metrograph project (Grant No. ANR-2011-NANO-004). It has been performed within the EMRP (European Metrology Research Program), project SIB51, Graphohm. The EMRP is jointly funded by the EMRP participating countries within EURAMET (European association of national metrology institutes) and the European Union.
\end{acknowledgments}
\appendix
\section{Weak localization measurements}
\begin{figure}[!ht]
\begin{center}
\includegraphics[width=8.5cm]{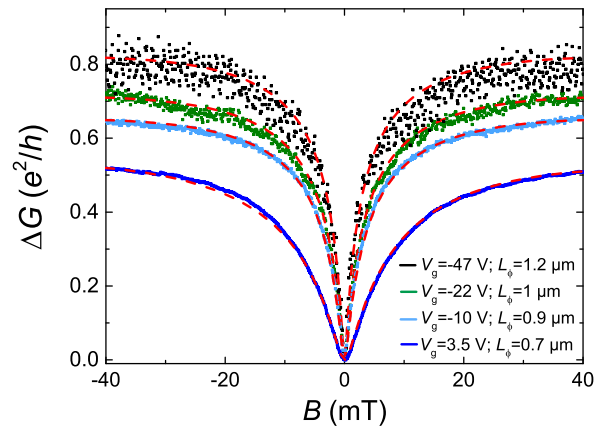}
\end{center}
\caption{Corrections to the conductance (one square) and fits (red dashed lines) with weak localization theory as a function of the magnetic field \emph{B} measured, at \emph{T}=0.3 K in sample S1, for several carrier density values: $n_\mathrm{s}\sim 0$ (at the CNP, $V_\mathrm{g}=3.5$ V, blue), $n_\mathrm{s}\sim -1.1\times10^{12}\mathrm{cm^{-2}}$ ($V_\mathrm{g}=-10$ V, light blue), $n_\mathrm{s}\sim -2.1\times10^{12}\mathrm{cm^{-2}}$ ($V_\mathrm{g}=-22$ V, green) and $n_\mathrm{s}\sim -3.9\times10^{12}\mathrm{cm^{-2}}$ ($V_\mathrm{g}=-47$ V, black). The values of the phase coherence length $L_\mathrm{\phi}$ deduced from fits by weak localization theory are indicated in figure.}
\end{figure}
Figure 7 shows the quantum corrections to the conductance as a function of the magnetic field, measured in sample S1 at \emph{T}=0.3 K and with a current \emph{I}=10 nA, for several carrier densities. Fitting these conductance curves with weak localization theory\cite{McCannWL2006}, one can deduce the phase coherence length $L_\mathrm{\phi}$, the inter-valley scattering length and the intra-valley scattering length. From the CNP to large hole carrier density $n_\mathrm{s}\sim -3.9\times10^{12}\mathrm{cm^{-2}}$, the phase coherence length $L_{\phi}$ varies from $\mathrm{0.7~\mu m}$ up to $\mathrm{1.2~\mu m}$.

\section{Numerical simulations}
In this section, we show some additional results to complement the main text.
\subsection{Local density of occupied states for given disorder and at different energies}
In fig. 6(c,d) of the main text we have shown the spatial distribution of the injected electrons at given energy and for two different levels of Anderson disorder along the line defect. In fig. 8, we illustrate a complementary simulation at $W$=2 eV and for injected electron energies $E=$ 100, 200 and 350 meV, corresponding to different localization regimes along the defect. We observe that the electrons injected from the right (source) contact flow along the bottom edge of the ribbon, as required by the spatial chirality of edge channels (electrons move along opposite directions at the two edges). Once the line defect reached, they can be transmitted to the drain contact along the same edge or backscattered along the top edge through the states of the defect. For $E$=100 meV, see fig. 8(a), the states along the line defect are localized and they cannot crosslink the edge channels. As a consequence, backscattering is not possible and the conductance is quantized to $2e^2/h$.
Note that a narrower ribbon may make the transmission of electrons through the localized states possible, thus allowing backscattering.
For $E$=200 meV and $E$=300 meV, see fig. 8(b,c), the states of the line defect are not localized enough to avoid transmission along the section of the ribbon, thus allowing for backscattering. As mentioned above, a wider ribbon width, i.e. a longer line defect length, would suppress electronic transmission from edge to edge and impede backscattering, thus restoring the conductance quantization as for $E=$100 meV. Note that the full scale in fig. 8(a-c) has been reduced to allow for the observation of the edge channels and the states around the line defect. However, a higher full scale highlights the presence of very localized states exactly on the atoms of the defect.
\begin{figure}[!ht]
\begin{center}
\includegraphics[width=8.5cm]{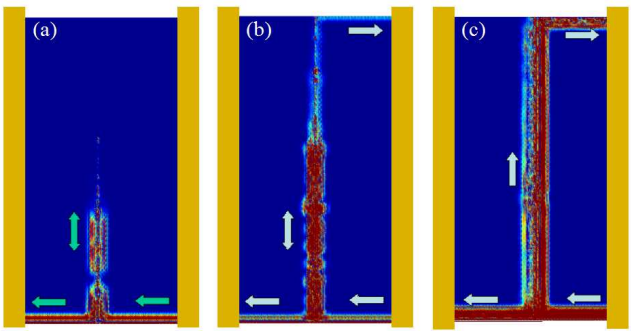}
\end{center}
\caption{Local density of injected electrons in the 200 nm wide ribbon with a 5-8 line defect and Anderson disorder of strength $W=$2 eV along the defect at energy 100 meV (a), 200 meV (b) and 300 meV (c).}
\end{figure}
\subsection{Dependence of the two-terminal conductance on magnetic field}
As indicated in the main text, we considered the joint effect of Anderson disorder along the line defect (with strength $W=1-4$ eV) and varying magnetic field (up to 120 T). The results are reported in fig. 9, where we scaled the energy as $E/\sqrt{B}$ in order to have the same position of the LLs for different fields and facilitate the comparison between different configurations.
The quality of the quantization increases with the magnetic field (especially at weak fields). This may be related to the fact that at higher magnetic field the magnetic length is shorter and then the states along the line defect are more confined in the region where disorder is, thus making them more sensitive to it. At high disorder strength and high magnetic field, very little backscattering is observed.
\begin{figure}[!ht]
\begin{center}
\includegraphics[width=8.5cm]{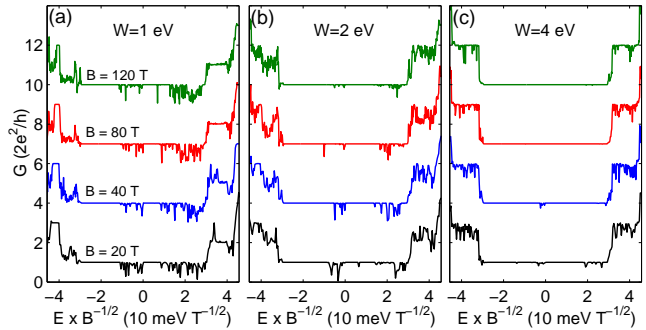}
\end{center}
\caption{Two-terminal magnetoconductance of a 200 nm wide graphene ribbon with a disordered 5-8 line defect under different magnetic fields. The strength of the disorder $W$ is 1 eV (a), 2 eV (b), 4 eV (c). The base lines were shifted for the sake of clarity.}
\end{figure}
\subsection{Origin of the nonchiral channels along the line defect}
In high magnetic field, extended states form along the line defect, which results in crosslinking opposite ribbon edge states. This can be qualitatively pictured by making a fictitious cut of the ribbon along the defect to obtain two uncoupled regions, where chiral edge states are generated for energy in between LLs, see fig. 10(a). Note that, in the region of the cut, the current flows in opposite direction in the two uncoupled ribbon parts (green and magenta arrows). When we join these two parts along the line defect, the counterpropagating edge states become spatially close the one another, see fig. 10(b). At this point, there are two possibilities, which depend both on the electron energy and the specific ribbon edge \cite{YAO_PRB88,Cummings}. We may have a gap along the weld joint, as, for example, in a perfect ribbon without any line defect. In this case, the counterpropagating edge channels cancel out, thus being unable to crosslink the ribbon edge channels. This is observed in fig. 6(a) of the main paper at energies $-250\lesssim E \lesssim -100$ meV, where the conductance is perfectly quantized. The second possibility is that the counterpropagating states survive and hybridize, thus giving rise to nonchiral edge states. This implies that electrons can flow in both directions. The level of spatial superposition of the channels determines the degree of their hybridization. For low hybridization degree, a residual chirality is expected, in the sense that electrons moving from the top edge to the bottom edge will be more concentrated at one side of the line defect, while electrons moving from the bottom edge to the top edge will be mainly located at the other side. However, due to the spatial proximity between the channels, a weak disorder is likely to induce a significant scattering between them. Indeed, as shown in the main paper, disorder is even able to localize these states, thus suppressing their extended nature.
\begin{figure}[!ht]
\begin{center}
\includegraphics[width=8.5cm]{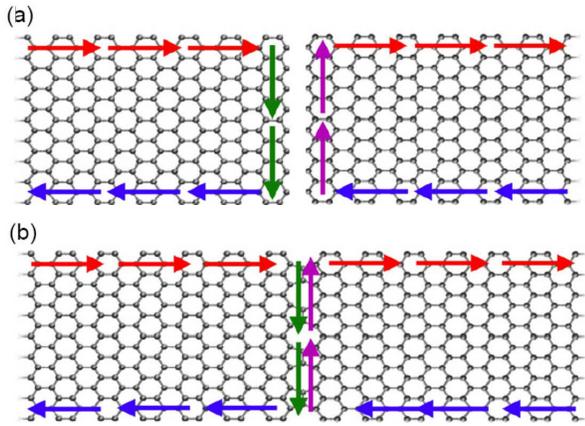}
\end{center}
\caption{(a) Chiral channels along the edges of the two uncoupled parts of a ribbon. (b) Channels for the complete ribbon with a line defect.}
\end{figure}

\providecommand{\noopsort}[1]{}\providecommand{\singleletter}[1]{#1}%

\end{document}